\documentclass[preprint,aps,prd,superscriptaddress,nofootinbib,tightenlines]{revtex4}

\usepackage{amsfonts}
\usepackage{multirow}
\usepackage{mathrsfs}
\usepackage{graphicx}
\usepackage{amsmath}
\usepackage{amssymb}
\usepackage{bm}
\usepackage{bbm}
\usepackage{color}
\usepackage{xcolor}
\usepackage{float}
\usepackage{slashed}
\usepackage{ulem}

\def\XXint#1#2#3{{\setbox0=\hbox{$#1{#2#3}{\int}$ }
\vcenter{\hbox{$#2#3$ }}\kern-.6\wd0}}

\newcommand{\nn}{\nonumber}

\newcommand{\beq}{\begin{equation}}
\newcommand{\eeq}{\end{equation}}
\newcommand{\bqa}{\begin{eqnarray}}
\newcommand{\eqa}{\end{eqnarray}}
\newcommand{\bseq}{\begin{subequations}}
\newcommand{\eseq}{\end{subequations}}

\begin{document}
\title{\mbox{}\\[10pt]
 Exclusive production of double light neutral mesons at the $e^+e^-$ colliders}

\author{Junliang Lu\footnote{lujl@ihep.ac.cn}}
\affiliation{School of Physics, Henan Normal University, Xinxiang 453007, China\vspace{0.2cm}}
\affiliation{Institute of High Energy Physics, Chinese Academy of Sciences, Beijing 100049, China\vspace{0.2cm}}

\author{Cai-Ping Jia\footnote{jiacp@ihep.ac.cn}}
\affiliation{School of Nuclear Science and Technology, Lanzhou University, Lanzhou 730000, China\vspace{0.2cm}}
\affiliation{Institute of High Energy Physics, Chinese Academy of Sciences, Beijing 100049, China\vspace{0.2cm}}

\author{Yu Jia\footnote{jiay@ihep.ac.cn}}
\affiliation{Institute of High Energy Physics,  Chinese Academy of Sciences, Beijing 100049, China\vspace{0.2cm}}
\affiliation{School of Physics, University of Chinese Academy of Sciences,
Beijing 100049, China\vspace{0.2cm}}

\author{Xiaonu Xiong\footnote{xnxiong@csu.edu.cn}}
\affiliation{School of Physics, Central South University, Changsha 410083, China\vspace{0.2cm}}

\date{\today}
\begin{abstract}
In this work we investigate the exclusive production of a pair of light neutral mesons in
$e^+e^-$ annihilation, where the final state bears an even $C$-parity. The production processes
can be initiated via the photon fragmentation or the non-fragmentation mechanism.
While the fragmentation contribution can be rigorously accounted,
the non-fragmentation contributions are calculated within the framework of collinear factorization,
where only the leading-twist light-cone distribution amplitudes (LCDAs)
of mesons are considered. Mediately solely by the non-fragmentation mechanism,
the production rates of double light neutral pseudoscalar mesons are too small to be observed
at the commissioning $e^+e^-$ facilities. In contrast, the production rates of a pair of light neutral vector mesons
are greatly amplified owing to the significant kinematic enhancement brought by the
fragmentation mechanism. It is found that, at $\sqrt{s}=3.77$ GeV, after including the
destructive interference between the non-fragmentation and
fragmentation contributions, the production rates for $e^+e^-\to \rho^{0}\rho^{0}$ and $\rho^0\omega$ can be lowered by about 10\% and 30\%
relative to the fragmentation predictions.
Future precise measurement of these exclusive double neutral vector meson production channels at {\tt BESIII} experiment
may provide useful constraints on the LCDAs of light vector mesons.
\end{abstract}

\maketitle
\section{Introduction}

The production of a pair of charged light mesons in $e^+e^-$ annihilation, such as $e^+e^-\to \pi^+\pi^-, \, K^+K^-,\, \rho^+\rho^-,\cdots$,
has been a widely studied topic from both experimental~\cite{CLEO:2005tiu,BaBar:2008fsh} and theoretical~\cite{Lepage:1979zb,Field:1981wx,Arbuzov:1997je,Lu:2006ut,Chen:2023byr} perspectives.
Since these processes proceed through the annihilation of $e^+e^-$ into a virtual photon, the final-state mesons pair must carry the odd $C$-parity.
These processes play a key role in determining the time-like electromagnetic form factors of charged mesons,
from which one can infer their internal partonic structures.

On the other hand, exclusive production of double neutral mesons in $e^+e^-$ annihilation, exemplified by $e^+ e^-\to \pi^0\pi^0, K_S K_S, \pi^0\eta, \eta \eta^\prime,
\rho^0\rho^0, \rho^0\omega, J/\psi J/\psi, J/\psi\rho^0, \cdots$, has also received considerable experimental and theoretical attention.
Since the final-state mesons bear an even $C$-parity, these processes must proceed via the annihilation of $e^+e^-$ into two virtual photons,
then followed by the transition from $\gamma^*\gamma^*$ to two neutral mesons. One naturally expects the production rate for such processes to be much more suppressed
with respect to that for production of a pair of charged mesons, due to the suppression brought by the extra powers of QED coupling constants.
In 2009, Kivel and Polyakov have considered the exclusive production of double light pseudoscalar mesons in collinear factorization~\cite{Kivel:2009xw}.
The cross section for $e^+e^-\to \pi^{0}\pi^{0}$ turns out to be several orders of magnitude smaller than that for 
$e^+e^-\to \pi^{+}\pi^{-}$.

In contrast to the $e^+e^-\to \pi^{0}\pi^{0}$ process,
the exclusive production of double neutral vector mesons is more interesting from a theoretical angle.
The $e^+e^-\to V_1 V_2$ processes
bear a distinct production mechanism: after annihilation of $e^+e^-$ into two virtual photons, two virtual photons can independently
fragment into two vector mesons.  In the non-fragmentation mechanism,  the valence quark and antiquark in each of the final-state mesons emerge from
the splitting of two different virtual photons, thus the virtuality of each photon propagator is of order $s$.
In contrast, the photon propagators in the two photon independent fragmentation processes
carry a typical virtuality of order $m_V^2$, thus the fragmentation contribution is enhanced by powers of $s/m_V^2$
with respect to the non-fragmentation contribution.
This kinematic enhancement factor may largely compensate for the suppression brought by extra powers of QED coupling $\alpha$, so that there is
good chance to observe these processes at $e^+e^-$ colliders.
Actually in 2008 {\tt BaBar} Collaboration reported the first observation of the $e^+e^-\to \rho^{0}\rho^{0}$ and $\rho^0\phi$ processes~\cite{BaBar:2006vxk}.
Shortly after, only taking into account the photon fragmentation contribution,
Davier {\it et al.}~\cite{Davier:2006fu} and Bodwin {\it et al.}~\cite{Bodwin:2006yd}
already found satisfactory agreement between their theoretical prediction and the measured value.

Although not being observed experimentally, there has been extensive theoretical study for $e^+e^-\to VV$ when the vector meson is a vector charmonium
such as $J/\psi$~\cite{Bodwin:2006yd,Gong:2008ce,Sang:2023liy,Huang:2023pmn}.
The non-fragmentation contribution can be accesses in the non-relativistic QCD (NRQCD) approach~\cite{Bodwin:1994jh}.
It is found that at $B$ factory, the interference between fragmentation and non-fragmentation amplitudes for double $J/\psi$ production
is destructive and non-negligible. For example, the leading-order NRQCD study indicates that including the interference effect can lower the
fragmentation contribution to the cross section by $30\%$~\cite{Bodwin:1994jh}.

To the best of our knowledge, the non-fragmentation contribution to exclusive production of a pair of light neutral vector mesons
has never been investigated in the literature.
This might be largely due to the perfect agreement between the predictions from the photon fragmentation mechanism~\cite{Davier:2006fu,Bodwin:2006yd}
and the {\tt BaBar} measurements for $\rho^0\rho^0$ and $\rho^0\phi$ production rates~\cite{BaBar:2006vxk},
which implies that the non-fragmentation contribution might be safely neglected for double neutral light vector meson production at $B$ factory energy.
However, one may expect that the non-fragmentation mechanism might yield a non-negligible contribution for
the double-$\rho^0$ production at {\tt BESIII} energy, since the importance of the non-fragmentation contribution increases with the decreasing center-of-mass energy.
Roughly speaking, since $(10.58\;{\rm GeV})^2/m^2_{J/\psi} \approx (3.77\;{\rm GeV})^2/m^2_{\rho}$, the non-negligible
non-fragmentation amplitude in $e^+e^-\to J/\psi J/\psi$ at $B$ factory suggests that the non-fragmentation contribution to
$e^+e^-\to \rho^0\rho^0$ at {\tt BESIII} experiment may also be non-negligible.

The goal of this work is to make a comprehensive analysis of exclusive production of double
light neutral vector mesons from $e^+e^-$ annihilation, including both fragmentation and non-fragmentation contributions.
We are especially interested in the {\tt BESIII} energy, which seems to be sensitive to the interference effect
induced by the non-fragmentation contribution. 
Following \cite{Kivel:2009xw}, we calculate the non-fragmentation amplitude in the framework of collinear factorization
tailored for hard exclusive reactions~\cite{Lepage:1980fj,Chernyak:1983ej}.
For the sake of completeness, we also revisit the exclusive production of a pair of light neutral pseudoscalar mesons~\cite{Kivel:2009xw}
and present a detailed phenomenological analysis.
Our numerical studies indicate that the production rates for $e^+e^-\to \rho^{0}\rho^{0},\rho^0\omega$ at $\sqrt{s}=3.770$ GeV are
large enough to be observed at the
commissioning {\tt BESIII} experiment.
It is rewarding to measure the production rate of this exclusive production channel with high accuracy,
so that one can clearly trace the footprint of the interference effect arising from the non-fragmentation mechanism.
By confronting the future measurement with our predictions, there is a good chance to constraint the
light-cone distribution amplitudes (LCDAs) of the $\rho$ and $\omega$ meson,
which might be useful for making more reliable predictions for the $B\to V$ form factor
and $B\to VV$ processes.

The rest of this paper is distributed as follows.
In Sec.~\ref{sec:amplitude:double:vector:meson}, we present
the amplitude for exclusive production of double neutral vector mesons from $e^+e^-$ annihilation,
including both fragmentation and non-fragmentation parts.
In Sec.~\ref{sec:amplitude:double:pseudoscalar:meson}, we recap the amplitude for exclusive production of a pair light neutral pseudoscalars from $e^+e^-$ annihilation,
which are governed by the non-fragmentation mechanism only. The effect of $\eta$-$\eta'$ mixing is also included.
In Sec.~\ref{Differential Cross Section}, we present the analytic expressions of the polarized and unpolarized cross sections for $e^+e^-\to V_1^0 V_2^0$ and $e^+e^-\to P_1^0 P_2^0$.
For the former, we also show the expressions of the interference and non-fragmentation parts.
In Sec.~\ref{Phenomenology}, we conduct a detailed numerical study for the angular distributions and the total cross sections
for a variety of double neutral meson production processes at {\tt BESIII} and {\tt Belle} energies. We pay special attention to
the $e^+e^-\to \rho^0\rho^0,\,\rho^0\omega$ channels at {\tt BESIII} experiment, in which the destructive interference effect becomes non-negligible.
It is advocated that the precise measurements of the angular distributions in these two channels
may offer novel means to constrain the LCDAs of the $\rho^0$ and $\omega$ mesons.
Finally we summarize in Sec.~\ref{Summary}.

\section{Production amplitudes of $e^+e^-\to V^0_1 V^0_2$}
\label{sec:amplitude:double:vector:meson}

\begin{figure}[htbp]
\begin{center}
\includegraphics[clip,width=0.7\textwidth]{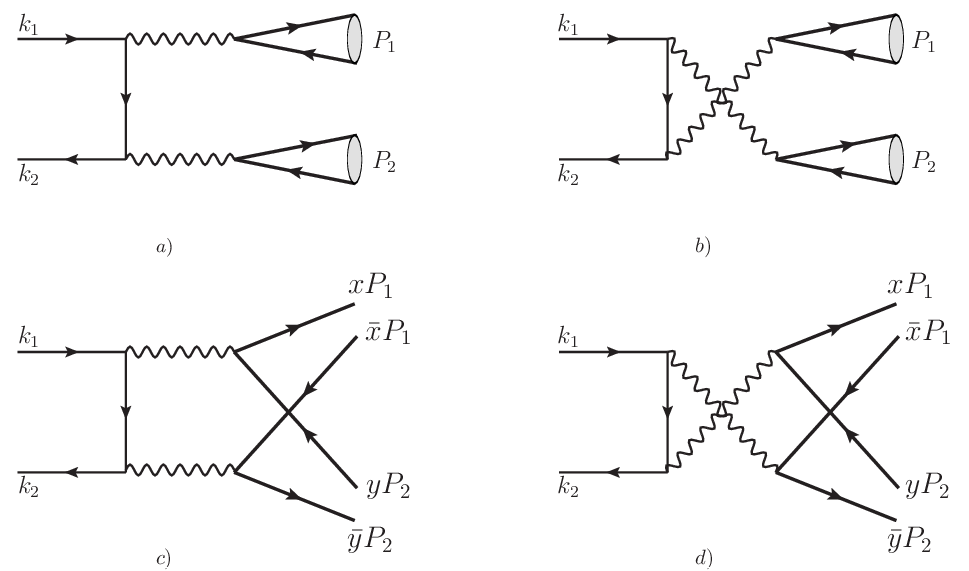}
\caption{The lowest-order Feynman diagrams for $e^+e^-$ annihilation into two neutral mesons.
The upper row represents the photon fragmentation contribution, while the lower row represents the non-fragmentation contribution.
 }
\label{ee:to:MM:Feynman:diagrams}
\end{center}
\end{figure}

We assume the $e^+e^-$  center-of-mass energy  $\sqrt{s}$ is high enough to warrant the applicability of perturbative QCD.
At the lowest order in QED and QCD coupling constants, there arise four Feynman diagrams that contribute to $e^+e^-\to V_1^0 V_2^0$,
which are depicted in Fig.~\ref{ee:to:MM:Feynman:diagrams}.
As stressed before, since the final-state mesons carry an overall even $C$ parity, this process has to be mediated with two-photon exchange.
Four Feynman diagrams can be categorized into the photon-fragmentation type (upper row) and photon non-fragmentation type (lower row).

In passing, we notice that our processes are similar, but much simpler than the high-energy production of double mesons
in $\gamma\gamma$ fusion, which has been widely studied both experimentally~\cite{Belle:2003xlt,Belle:2004bpk,Belle:2007ebm} and
theoretically~\cite{Brodsky:1981rp,Benayoun:1989ng,Chernyak:2006ms,Chernyak:2006dk,Chernyak:2014wra,Klusek:2009yi}.
The important difference is that, a hard gluon has to be exchanged between the outgoing quarks and antiquarks for the latter,
while it is not necessary in our case, at least at the lowest order in $\alpha_s$.

\subsection{Photon fragmentation mechanism}

As indicated in Fig.~\ref{ee:to:MM:Feynman:diagrams}$a)$ and $b)$, two neutral mesons can be produced through two photon independent fragmentation mechanisms.
Since the $C$-parity conservation forbids a photon to convert into a pseudoscalar meson, so we only need to consider the fragmentation production of
two neutral vector mesons. Let us denote the momenta of the incoming $e^-$ and $e^+$ by $k_1$ and $k_2$, and denote the momenta of the outgoing meson pairs by $P_1$ and $P_2$.
The momenta are subject to the constraints $k_1^2=k_2^2=0$, $k_1\cdot k_2 =s/2$, and $P_i^2=m^2_{V_i}$ ($i=1,2$),
where $\sqrt{s}$ signifies the center-of-mass energy, $m_{V_i}$ indicating the mass of the $i$-th neutral vector meson.
The production amplitude from the two-photon independent fragmentation mechanism reads~\footnote{
Note that the photon-to-vector meson fragmentation mechanism is often referred to as the vector meson dominance (VMD) model~\cite{Sakurai:1960ju} in literature
(for example, see Ref.~\cite{Davier:2006fu}).
Nevertheless, in our process the vector mesons manifest themselves as the asymptotic out states,
rather than appear in the intermediate state (vector meson propagator),
therefore there is no model dependence in our treatment of the fragmentation amplitude.}
\bqa
\mathcal{A}^{\mathrm{fr}}(e^+e^-\to V_1^0(\lambda_1) V_2^0(\lambda_2)) &= &
-{e^2  g_{V_1} g_{V_2} \over m^2_{V_1}  m^2_{V_2}}  {\bar{v}}(k_{2})\Big[\frac{{\not \!
\varepsilon}_{2}^{*}(\lambda_2)({\not\! k}_{1}-{\not \! P}_{1}){\not \! \varepsilon}_{1}^{*}(\lambda_1)}{(k_{1}-P_{1})^{2}}
\nn\\
&+ & {{\not\!\varepsilon}_{1}^{*}(\lambda_1)({\not \! k}_{1}-{\not \! P}_{2}){\not \!\varepsilon}_{2}^{*}(\lambda_2)\over
(k_{1}-P_2)^2 } \Big] u(k_{1}),
\label{Frag:ampl:double:vector:meson}
\eqa
where $\varepsilon_{1,2}$ signify the polarization vectors of two outgoing vector mesons, and $g_V$ encodes the effective coupling between the
photon and neutral vector meson~\cite{Bodwin:2006yd}:
\beq
\langle V^0(P,\lambda)|j^\mu_{\rm EM}|\,0\,\rangle = g_V \varepsilon^{(\lambda)\mu\,*}(P).
\label{fv}
\eeq
The electromagnetic current $j^\mu_{\rm EM}$ can be decomposed into the isospin basis,
\bqa
 j^\mu_{\rm EM} &= & e_u  \bar{u}\gamma^\mu u+ e_d  \bar{d}\gamma^\mu d+ e_s  \bar{s}\gamma^\mu s+\cdots
\nn\\
 &= &
 {e_u-e_d\over 2}  \bar n \sigma^3 \gamma^\mu n \Big|_{I=1}   +  {e_u+e_d\over 2}\bar n {\tt 1}\gamma^\mu n \Big |_{I=0} +
 e_s  \bar{s}\gamma^\mu s \Big |_{I=0}+\cdots,
\label{EM:current}
\eqa
where $n=(u\;\;d)^T$ denotes the quark isospin doublet in the first generation,
$\sigma^3$ and $\tt 1$ signify the third Pauli matrix and the unit matrix in the isospin space.
The subscripts in the second line of \eqref{EM:current}
are reminiscent of the isospin carried by various quark vector currents.  $e_u={2\over 3}$, $e_d=e_s=-{1\over 3}$ indicate the electric charges of the light quarks.

It is a common practice to work with the decay constant of a light neutral vector mesons, $f_V$.
For three species of lowest-lying neutral vector mesons, the decay constants are defined via
\bseq
\bqa
&& \langle \rho^b(P,\lambda)| \bar{n} \sigma^a \gamma_\mu n| 0\rangle = \delta^{ab} \sqrt{2}   f_\rho \,m_\rho  \varepsilon^{(\lambda)*}_\mu(P),
\qquad (a,b=1,2,3)
\label{rho0:decay:constant}\\
&& \langle \omega(P,\lambda)| \bar n  {\tt 1} \gamma_\mu n |0\rangle =  \sqrt{2} f_\omega m_\omega  \varepsilon^{(\lambda)*}_\mu(P),
\\
&& \langle \phi(P,\lambda)|  \bar s \gamma_\mu s |0\rangle = f_\phi m_\phi  \varepsilon^{(\lambda)*}_\mu(P),
\eqa
\label{def:neutral:vector:meson:decay:constants}
\eseq
with $\rho^3$ identified with $\rho^0$ in \eqref{rho0:decay:constant}. $m_V$ signifies the mass of the vector meson.

The photon-to-$V$ coupling $g_V$ is related to the vector meson decay constant via
\beq
g_V = {\cal Q}_V f_{V} m_{V}
\label{Relation:gV:fV}
\eeq
for $V=\rho^0,\omega,\phi$.
Here ${\cal Q}_V$ characterizes the effective electric charge of quark affiliated with each neutral vector meson:
\beq
{\cal Q}_{\rho^0} = {1 \over \sqrt{2} } (e_u-e_d)={1\over \sqrt{2}}, \qquad  {\cal Q}_\omega = {1\over \sqrt{2}}(e_u+e_d) = {1\over 3\sqrt{2}},  \qquad
{\cal Q}_\phi=e_s=-{1\over 3}.
\label{effective:quark:charge}
\eeq
The occurrence of the ${\cal Q}_V$ factor stems from the fact that the $\rho^0$ and $\omega$ mesons contain both $u$ and $d$ quarks in their flavor wave function, {\it i.e.},
$\rho^0={1\over \sqrt{2}} |u\bar{u}-d\bar{d}\rangle$, $\omega={1\over \sqrt{2}} |u\bar{u}+d\bar{d}\rangle$.

The leptonic width of the neutral vector meson then becomes
\beq
\Gamma(V^0\to e^+e^-) ={4\pi \alpha^2\over 3} {g_V^2\over m^3_V} = {4\pi {\cal Q}_V^2 \alpha^2\over 3} {f_V^2\over m_V}.
\label{Leptonic:width:neutral:vector:meson}
\eeq
One thus can deduce the values of $g_V$ and $f_V$ from the precisely measured leptonic width.

\subsection{Non-Fragmentation contribution}

The lowest-order non-fragmentation diagrams are depicted in Fig.~\ref{ee:to:MM:Feynman:diagrams}$c)$ and $d)$, where the quark and anti-quark
in each outgoing vector meson stem from the splitting of two different photons. The photon propagators bear a typical
virtuality of order $s$, much greater than $m_V^2$ in the fragmentation mechanism, therefore we expect such non-fragmentation contribution
to be suppressed with respect to the fragmentation contribution.
For a hard exclusive reaction involving hadrons, the amplitude can be expressed as a convolution between the perturbatively-calculable hard scattering kernel
and the nonperturbative yet universal LCDAs of hadrons.
Following Ref.~\cite{Kivel:2009xw}, we apply the collinear factorization to investigate the non-fragmentation contribution to $e^+e^-\to V_1^0 V_2^0$,
at the lowest order in $1/s$ and $\alpha_s$.

As the key input in collinear factorization, LCDAs characterize the momentum distribution of the valence quarks inside a fast-moving hadron.
The LCDAs of a light vector mesons carrying helicity $\lambda$
are related to the following quark correlator with light-like separation through Fourier transform~\cite{Beneke:2000wa}~\footnote{For simplicity, here we assume that
$V$ is a neutral vector meson made of a single flavor, {\it e.g.}, the $\phi$ meson. The
LCDAs of $\rho^0$ and $\omega$ can be obtained by inserting the matrices $\sigma^3$ or ${\tt 1}$ in the isospin space.}:
\bqa
& &\langle V(P,\lambda)|{\bar q}_{\alpha}(u_2)[u_2,u_1]q_{\beta}(u_1)|\,0\,\rangle
\nn\\
  & & = -\frac{i}{4}\int_0^1 \!\! dx e^{i(xp \cdot u_2 + \bar{x} p \cdot u_1)}\left\{f_V m_V p_\mu\frac{\varepsilon_\parallel^*\cdot u}{p\cdot u}\phi_\parallel(x)
    +f_\perp{\not \varepsilon_\perp^*}\not\!{p}\phi_\perp(x)+\cdots\right\}_{\beta\alpha},
\label{Def:LCDA:vector:meson}
\eqa
with $u\equiv u_1-u_2$ and $u^2=0$. $p_\mu$ is a light-like four-momentum which is related to the physical meson momentum $P_\mu$ via
$p_\mu =P_\mu - m_V^2 u_\mu /(2P\cdot u)$.
$\varepsilon_\parallel$ and $\varepsilon_\perp$ denote the polarization vector of the vector meson with helicity $\lambda$, and
$\phi_{\parallel}(x)$ and $\phi_\perp(x)$ represent the leading-twist LCDAs for a longitudinally- and transversely-polarized vector meson,
with $x$  ($\bar{x}\equiv 1-x$) sifnifying the light-cone momentum fraction carried by the quark (anti-quark).
The ellipses represent all the neglected higher-twist contributions, which are irrelevant to our purpose.
$[u_2,u_1]$ in \eqref{Def:LCDA:vector:meson} signifies the gauge link,
\begin{equation}
    [u_2,u_1]=\mathcal{P}\left\{\text{exp}\left[-ig_s\int_0^1dt\, u_\mu A^\mu(tu_2+(1-t)u_1)\right]\right\},
\end{equation}
which is inserted to ensure the gauge invariance of the LCDAs of a vector meson.

In addition to the usual decay constant $f_V$ in \eqref{def:neutral:vector:meson:decay:constants}, a new decay constant $f_{V\,\perp}$ also arises in
\eqref{Def:LCDA:vector:meson}, which is related to the transversely-polarized vector meson:
\bseq
\bqa
&& \langle \rho^b(P,\lambda)| \bar n \sigma^a \sigma_{\mu\nu}  n|0\rangle = -\delta^{ab} \sqrt{2} 
f_{\rho \perp}(P_{\mu} \varepsilon^{*}_{\nu}(\lambda)-
P_{\nu}\varepsilon^{*}_{\mu}(\lambda)),
\qquad (a,b=1,2,3)
\label{rho0:decay:constant:transverse}\\
&&  \langle \omega(P,\lambda)| \bar n \sigma_{\mu\nu} n|0\rangle = - \sqrt{2} 
f_{\omega \perp}(P_{\mu} \varepsilon^{*}_{\nu}(\lambda)-
P_{\nu}\varepsilon^{*}_{\mu}(\lambda)),
\\
& &\langle \phi(P,\lambda)| \bar s \sigma_{\mu\nu}  s|0\rangle = - f_{\phi \perp}
(P_{\mu} \varepsilon^{*}_{\nu}(\lambda)-
P_{\nu}\varepsilon^{*}_{\mu}(\lambda)).
\eqa
\label{def:neutral:vector:meson:decay:constants:transverse}
\eseq

To expedite computing the vector meson exclusive production amplitude,
it is convenient to apply the light-cone projectors in the momentum space~\cite{Beneke:2000wa}:
\beq
   M_{\beta\alpha}^{V}(x)= M_{\parallel\beta\alpha}^{V}(x)+ M_{\perp \beta \alpha}^{V}(x),
\eeq
where
\bseq
\bqa
M_{||}^{V}(x) &=&-\frac{if_{V}}{4}\frac{m_{V}(\varepsilon^* \cdot n_{+})}{2E} E{\not\! n}_{-}\phi_{\parallel}(x)+\cdots,
\\
M_{\perp}^{V}(x) &=&-\frac{if_{\perp}}{4}E{\not \!\varepsilon^*_{\perp}}{\not \! n}_{-}\phi_{\perp}(x)+\cdots
\eqa
\label{light-cone:projectors:parallel:perp}
\eseq
are projectors for the longitudinally- and transversely-polarized vector meson, respectively.
A pair of conjugate light-like four-vectors $n_\pm$ are introduced through $P^\mu = E n_-^\mu+ m_V^2  n_+^\mu/(4E)$.
Since we are not interested in the power-suppressed high-twist corrections, suffices it that to retain only the leading-twist LCDAs in
\eqref{light-cone:projectors:parallel:perp}, and it is legitimate to
approximate the momentum of the vector meson by $P^\mu\approx p^\mu =  E n_-^\mu$.

Substituting  \eqref{light-cone:projectors:parallel:perp} into the quark amplitude $e^+e^-\to q_1(x p_1) \bar{q}_1(y p_2) + q_2(\bar{y} p_2) \bar{q}_2(\bar{x} p_1)$,
the nonfragmentation amplitudes of double neutral vector meson production can be obtained through
\bqa
&& \mathcal{A}^{\mathrm{nfr}}(e^+e^-\to V_1^0(\lambda_1) V_2^0(\lambda_2)) = e^{4} \kappa_{V_1 V_2} \int\!\!\!\int\!\! dx dy \; {\text{Tr}[M_{\lambda_1}^{V_1}(y)\gamma^{\beta}M_{\lambda_2)}^{V_2}(x)\gamma^{\alpha}]\over (xp_{1}+yp_{2})^2({\bar x}p_{1}+{\bar y}p_{2})^2}
 \nn\\
 &\times&
 \overline{v}(k_2) \Big[\frac{\gamma_{\beta}({\not k}_{1}-x{\not\! p}_{1}-y{\not\! p}_{2})\gamma_{\alpha}}{(k_{1}-xp_{1}-yp_{2})^{2}}
 +\frac{\gamma_{\alpha}({\not \! k}_{1}-{\bar x}{\not\! p}_{1}-{\bar y}{\not\! p}_{2})\gamma_{\beta}}{(k_{1}-{\bar x}p_{1}-{\bar y}p_{2})^{2}}\Big]u(k_{1}).
\label{Non:fragmentation:ampl:double:vector:mesons}
\eqa
The factor $\kappa_{V_1 V_2}$ represents the squared effective quark electric charge, which arises from the superposition of the quark amplitude in the flavor space.
The concrete values of $\kappa_{V_1 V_2}$ are tabulated in Table~\ref{tab:charge}~\footnote{Note we
have omitted the small $\rho^0$-$\omega$ and $\omega$-$\phi$ mixing effects for simplicity.}.

\begin{table}[h]
    \centering
    \begin{tabular}{|c|c|c|c|c|c|} \hline
         channel &  $\kappa_{V_1 V_2}$& value & channel & $ \kappa_{P_1 P_2}$&value\\ \hline
         $ \rho^0 \rho^0$&  $\frac{1}{2}(e_u^2+e_d^2)$& $\frac{5}{18}$ & $\pi^0 \pi^0$& $\frac{1}{2}(e_u^2+e_d^2)$&$\frac{5}{18}$\\ \hline
         $\rho^0 \omega$&  $\frac{1}{2}(e_u^2-e_d^2)$& $\frac{1}{6}$ & $ \pi^0 \eta_q$& $\frac{1}{2}(e_u^2-e_d^2)$ &$\frac{1}{6}$\\ \hline
         $ \omega \omega$&  $\frac{1}{2}(e_u^2+e_d^2)$& $\frac{5}{18}$ & $ \eta_q \eta_q$& $\frac{1}{2}(e_u^2+e_d^2)$&$\frac{5}{18}$\\ \hline
         $ \phi \phi$&  $e_s^2$& $\frac{1}{9}$ & $ \eta_s \eta_s$& $e_s^2$&$\frac{1}{9}$\\ \hline
 & & & $K_0{\bar K}_0$& $e_d e_s$&$\frac{1}{9}$\\\hline
    \end{tabular}
\caption{The squared effective quark charge for $e^-e^+\to M_1 M_2$ in the non-fragmentation amplitude.}
\label{tab:charge}
\end{table}

After some straightforward manipulation, we obtain the non-fragmentation amplitudes from various helicity configurations:
\bseq
\begin{align}
& \mathcal{A}^{\mathrm{nf}}_{\pm 1,\mp 1}
 =
 \frac{e^{4} \kappa_{V_1V_2} f_{V_1\perp} f_{V_2\perp}}{N_c s^2}  {\bar v}(k_2)
[ (\varepsilon^*_{1\perp} \cdot k_1){\not \varepsilon_{2\perp}^*}+(\varepsilon^*_{2\perp} \cdot k_1){\not \!\varepsilon_{1\perp}^*}]
 u(k_1)\, G_{\pm1,\mp1}(\cos\theta),
\\
&\mathcal{A}^{\mathrm{nf}}_{0,0}
 =
 \frac{e^{4}  \kappa_{V_1V_2} f_{V_1}f_{V_2}}{N_c s^2}
{\bar v}(k_2)({\not\! p}_1-{\not\! p}_2)
u(k_1) \, G_{0,0}(\cos\theta),
\\
& \mathcal{A}^{\mathrm{nf}}_{\pm 1,\pm 1}
 =\mathcal{A}^{\mathrm{nf}}_{\pm 1,0}
 =\mathcal{A}^{\mathrm{nf}}_{0,\mp 1} =0.
\end{align}
\label{helicity:amplitude:nonfrag:double:vector:meson}
\eseq
where
\bseq
\bqa
& & G_{\pm1,\mp1}(z) = \int\!\!\!\int\!\! dx dy \frac{\phi_{1,\perp}(x)}{x{\bar x}}
\frac{\phi_{2,\perp}(y)}{y{\bar y}}
{x+y-2xy\over (x+y-2xy)^2-z^2 (x-y)^2},
\\
&& G_{0,0}(z) =  z \int\!\!\!\int\!\! dx dy \frac{\phi_{1,\parallel}(x)}{x{\bar x}}
    \frac{\phi_{2,\parallel}(y)}{y{\bar y}}
    \frac{x{\bar x}+y{\bar y}}{(x+y-2xy)^2-z^2 (x-y)^2}.
\eqa
\label{Auxiliary:Angular:functions:different:helcity}
\eseq
It is reassuring that the two-fold integration in $G(z)$ renders a finite result,
provided that the leading-twist LCDAs assume the standard endpoint behavior: $\phi_\parallel(x)\sim \phi_\perp(x)\propto x$.

We also remark that, $\mathcal{A}^{\mathrm{nf}}_{0,0}$ in \eqref{helicity:amplitude:nonfrag:double:vector:meson} is
identical to the amplitude for producing double pseudoscalar mesons in \eqref{amplitude:double:pseudoscalar:meson},
except the corresponding replacement with the leading-twist LCDA and the decay constant is made.
This is not a coincidence, since the hard-scattering amplitude in collinear factorization is only sensitive to the meson helicity,
rather than the meson's spin.

\section{Production amplitudes of $e^+e^-\to P^0_1 P^0_2$}
\label{sec:amplitude:double:pseudoscalar:meson}

For the sake of completeness, in this section we recapitulate the calculation of exclusive production of a pair of light neutral pseudoscalar mesons from $e^+e^-$ collision.
We also consider the production channels involving  $\eta(\eta')$ and $K_S$ in addition to the $e^+e^-\to \pi^{0}\pi^{0}$ process
investigated by Kivel and Polyakov~\cite{Kivel:2009xw}.
Due to the mismatch of the quantum number between photon and pseudoscalar meson, the photon mechanism is absent in these processes, and we only
need take into account the non-fragmentation contributions as indicated by Fig.~\ref{ee:to:MM:Feynman:diagrams}$c)$ and $d)$.

Analogous to the LCDAs of a vector meson as introduced in \eqref{Def:LCDA:vector:meson},
the leading-twist LCDA of a pseudoscalar meson
is related to following light-cone quark correlator through Fourier transform~\cite{Beneke:2000wa}
\begin{align}
\langle P(P)|{\bar q}_{\alpha}(u_2)[u_2,u_1]q_{\beta}(u_1)|\,0\,\rangle=&\frac{if_P}{4}\int_0^1dx e^{i(xp\cdot u_2+\bar{x}p\cdot u_1)}\left\{\not\!{p}\gamma_5\phi(x)+\cdots\right\}_{\beta\alpha},
\end{align}
where the notations are identical to \eqref{Def:LCDA:vector:meson}.  Since the pseudoscalar merely carries zero helicity,
there arises only one leading-twist LCDA $\phi(x)$.
$f_P$ signifies the decay constant of a pseudoscalar meson, which is defined through
\bseq
\bqa
&& \langle \pi^b(P)| \bar{n} \sigma^a \gamma^\mu\gamma_5 n| 0\rangle = i \delta^{ab} \sqrt{2}   f_\pi P^\mu,
\qquad (a,b=1,2,3)
\label{pion:decay:constant}\\
&& \langle \eta_q (P)| \bar{n}  {\tt 1} \gamma^\mu\gamma_5 n |0\rangle =  i \sqrt{2} f_{\eta_q} P^\mu,\qquad \langle \eta_s (P)| \bar{s} \gamma^\mu\gamma_5 s |0\rangle =  i f_{\eta_s} P^\mu,
\\
&& \langle K^0(P) \vert  \bar{d} \gamma^\mu\gamma_5 s |0\rangle = i f_K  P^\mu.
\eqa
\label{def:neutral:pseudoscalar:meson:decay:constants}
\eseq
with $\pi^3$ identified with $\pi^0$ in \eqref{pion:decay:constant}. 

The light-cone projector in momentum space is particularly simple for a pseudoscalar~\cite{Beneke:2000wa}:
\beq
M_{\beta\alpha}^{P}(x) = i\frac{if_{P}}{4}
 \left({\not\! p}\gamma_{5}\right)_{\beta\alpha}\,\phi(x),
\label{light-cone:projectors:pseudoscalar:leading:twist}
\eeq
where $p^\mu = E n_-^\mu$ is a light-like momentum, the same as in \eqref{light-cone:projectors:parallel:perp}.

Substituting  \eqref{light-cone:projectors:pseudoscalar:leading:twist} into the quark amplitude $e^+e^-\to q_1(x p_1) \bar{q}_1(y p_2) + q_2(\bar{y} p_2) \bar{q}_2(\bar{x} p_1)$,
one then obtains the intended amplitude for $e^+e^-\to P_1^0 P_2^0$. This can be achieved by simply replacing the light-cone projectors
$M_{\parallel(\perp)}^V$ with $M^P$ in \eqref{Non:fragmentation:ampl:double:vector:mesons}, and the resulting amplitude is
\beq
 \mathcal{A}(e^+e^-\to P_1^0 P_2^0 ) =
 {e^{4} \kappa_{P_1 P_2} f_{P_1}f_{P_2}\over N_c s^2}\, {\bar v}(k_2)({\not \! p}_1-{\not\! p}_2) u(k_1) \, G_{P,P}(\cos\theta),
\label{amplitude:double:pseudoscalar:meson}
\eeq
with
\begin{align}
    G_{P,P}(z) = z \int\!\!\!\int\!\!  dx dy \frac{\phi_{1}(x)}{x{\bar x}}
    \frac{\phi_{2}(y)}{y{\bar y}}
    \frac{x{\bar x}+y{\bar y}}{(x+y-2xy)^2- z^2 (x-y)^2}.
\label{Auxiliary:Angular:functions:double:pseudoscalar}
\end{align}
These expressions are in agreement with (2.1) and (2.2) of Ref.~\cite{Kivel:2009xw}.  Again the angular function $G_{P,P}(\cos\theta)$ is finite as long as
$\phi_{i}(x)\propto x$ near the endpoint regime.

The factor $\kappa_{P_1 P_2}$ in \eqref{amplitude:double:pseudoscalar:meson} has the same physical meaning as
$\kappa_{V_1 V_2}$ in \eqref{Non:fragmentation:ampl:double:vector:mesons}.
It represents the squared effective quark electric charge, which emerges from the superposition of the quark amplitude in the flavor space.
The concrete values of various $\kappa_{P_1 P_2}$ have been tabulated in Table~\ref{tab:charge}.

Some special care is needed when the final-state mesons involve $\eta$ or $\eta'$.  The pseudoscalar states appearing in \eqref{amplitude:double:pseudoscalar:meson}
are actually not the physical $\eta$ and $\eta^\prime$, but two iso-singlets $\eta_{q}=(u{\bar u}+d{\bar d})/\sqrt{2}$ and $\eta_{s}=s{\bar s}$~\footnote{Note that in order to
predict $e^+e^- \to \pi^0 \eta(\eta')$ production rate, we only need consider the amplitude of $e^+e^- \to \pi^0 \eta_q$, since
there arises no such reaction as $e^+e^- \to \pi^0 \eta_s$ from the non-fragmentation mechanism.}.
The physical $\eta$ and $\eta'$ states are the mixtures between $\eta_q$ and $\eta_s$~\cite{Feldmann:1998vh}~\footnote{In literature there has been speculation that
a pseudoscalar glueball may also mix with $\eta$ and $\eta'$ due to chiral anomaly~\cite{Cheng:2008ss}. Since the photon does not directly couple
to gluon, we will not dwell on this complication for exclusive $\eta$ and $\eta'$ production.}:
\begin{eqnarray}
   \begin{pmatrix}
\eta \\
\eta' \\
   \end{pmatrix}
   =
    \begin{pmatrix}
\cos{\phi} & -\sin{\phi} \\
\sin{\phi} & \cos{\phi} \\
   \end{pmatrix}
     \begin{pmatrix}
\eta_{q} \\
\eta_{s} \\
   \end{pmatrix},
\end{eqnarray}
The mixing angle has been determined to be $\phi=(38.8\pm2.4)^\circ$~\cite{Ottnad:2017bjt}.

Some explanation on the kaon pair exclusive production is in order. It is experimentally favorable to look for the $e^+e^-\to K_S K_S$ process.
However, the kaon pair appearing in \eqref{amplitude:double:pseudoscalar:meson} are the flavor eigenstates $K_0$ and $\overline{K}_0$.
Since the CP violation in neutral kaon is quite small, we can approximately assume $|K^0_S\rangle= {1\over \sqrt{2}}(|K^0\rangle+|\overline{K}^0\rangle)$.
Consequently, the amplitudes for $e^+e^-\to K^0_S K^0_S$ and $e^+e^-\to K^0 \overline{K}^0$ turn out to be identical.

\section{Differential Cross Sections for double neutral meson production}
\label{Differential Cross Section}

In the preceding sections we have obtained the amplitudes of exclusive neutral meson pair production, and
it is straightforward to deduce the corresponding differential polarized cross sections:
\beq
{d\sigma_{\lambda_1,\lambda_2}(e^+ e^- \to M_1(\lambda_1) M_2(\lambda_2))\over d\cos\theta}
= {1\over 2s}\frac{1}{16\pi}\frac{2|{\bf P}_1|}{\sqrt{s}}\frac{1}{4} \sum_{\text{spin}}|{\cal A}|^2,
\eeq
where $\lambda_i$ ($i=1,2$) signifies the helicity of the $i$-th meson in the final state, and the spins of
electron and positron spin have been averaged over.
$\theta$ represents the polar angle between the first meson's three-momentum and the moving direction of the electron beam.
$|{\bf P}_1|$ denotes the magnitude of the meson three-momentum in the center-of-mass frame,
$|{\bf P}_1|=\lambda^{\frac{1}{2}}(s,m_{V_1}^2,m_{V_2}^2)/(2\sqrt{s})$, with
\beq
\lambda(x,y,z)=x^2+y^2+z^2-2xy-2xz-2yz.
\eeq

\subsection{Production rates for double neutral vector mesons}

As elucidated in Sec.~\ref{sec:amplitude:double:vector:meson}, for double neutral vector meson production,
both the fragmentation and non-fragmentation contributions have been considered. The differential cross sections
can thus be decomposed into three pieces:
\bqa
&& {d\sigma_{\lambda_1,\lambda_2}(e^+ e^- \to V^0_1(\lambda_1) V^0_2(\lambda_2))\over d\cos\theta} =  \frac{1}{2s}\frac{1}{16\pi}\frac{2|{\bf P}_1|}{\sqrt{s}}\frac{1}{4}\sum_{\text{spins}}\left|\mathcal{A}^\text{fr}+\mathcal{A}^\text{nf}\right|^2,
\nn\\
&& = \frac{d\sigma^{\rm fr}_{\lambda_1,\lambda_2}}{d\cos\theta} + \frac{d\sigma^{\rm int}_{\lambda_1,\lambda_2}}{d\cos\theta}+
\frac{d\sigma^{\rm nfr}_{\lambda_1,\lambda_2}}{d\cos\theta}.
\eqa
At the commissioning $e^+e^-$ collision experiments,
the dominant contribution comes from the fragmentation part, and the next important contribution comes from
the interference part. The non-fragmentation part is anticipated to yield a negligible contribution.

\subsubsection{Fragmentation part}\label{fragmentation}

The photon fragmentation contribution to the polarized cross sections for double neutral vector meson production was first obtained
by Bodwin {\it et al.} in 2006~\cite{Bodwin:2006yd}. We have confirmed the correctness of their expressions.
Here for the sake of completeness, we enumerate the explicit expressions of the polarized cross section for all possible helicity
configurations of the final-state vector mesons.

From \eqref{Frag:ampl:double:vector:meson} one can readily deduce the fragmentation contribution to the
differential cross section for each helicity configurations $(\lambda_1,\lambda_2)$~\cite{Bodwin:2006yd}:
\beq
\frac{d\sigma^\text{fr}_{\lambda_1,\lambda_2}}{d\cos{\theta}}
   =\frac{16\pi^3\alpha^4 {\cal Q}_{V_1}^2 {\cal Q}_{V_2}^2f_{V_1}^2f_{V_2}^2r_{V_1}^2r_{V_2}^2\lambda^{1/2}(1,r_{V_1}^{2},r_{V_2}^2)
   F_{\lambda_1,\lambda_2}(r^2_{V_1},r^2_{V_2},\cos{\theta})}{s^3r_{V_1}^4r_{V_2}^4 [\lambda(1,r_{V_1}^2,r_{V_2}^2)\sin^2{\theta} + 4r_{V_1}^2r_{V_2}^2]^2},
\label{polarized:differential:cross:section:vector}
\eeq
with the dimensionless ratios $r_{V_i}\equiv m_{V_i}/\sqrt{s}$ ($i=1,2$).
The helicity-dependent functions $F_{\lambda_1,\lambda_2}$ read
\bseq
\bqa
&& F_{\pm1,\mp1}(r^2_{V_1},r^2_{V_2},z)=(1-z^4)(1-r^2_{V_1}-r^2_{V_2})^2,
\\
&&  F_{\pm1,0}(r^2_{V_1},r^2_{V_2},z)=r^2_{V_2}[2r^4_{V_1}(z^4+6z^2+1)-4r^2_{V_1}(1-r^2_{V_2})(1-z^4)
\\ \nn
&&\qquad\qquad\qquad\qquad+2(1-r^2_{V_2})^2(z^2-1)^2]
  ,
\\
 && F_{0,\pm1}(r^2_{V_1},r^2_{V_2},z)=F_{\pm1,0}(r^2_{V_2},r^2_{V_1},z)
 ,
\\
&&  F_{\pm1,\pm1}(r^2_{V_1},r^2_{V_2},z)=4r^2_{V_1}r^2_{V_2}z^2(1-(r^2_{V_1}-r^2_{V_2})^2)-z^4(r^4_{V_1}+r^4_{V_2}
\\ \nn
&&\qquad\qquad\qquad\qquad-r^2_{V_1}(2r^2_{V_2}+1)-r^2_{V_2})^2+(r^2_{V_1}-r^2_{V_2}-r^4_{V_1}+r^4_{V_2})^2
,
\\
&&  F_{0,0}(r^2_{V_1},r^2_{V_2},z)=16r_{V_1}^2r_{V_2}^2 z^2(1-z^2),
\eqa
\eseq

The asymptotical behaviors of various polarized cross sections in \eqref{polarized:differential:cross:section:vector} in the high energy limit
are
\bseq
\bqa
& & {d\sigma^\text{fr}_{\pm_1,\mp_1}\over d\cos{\theta}} \sim  {\alpha^4 \over s},
\label{Fragmentation:cross:section:Leading:helicity:scaling}
\\
& & {d\sigma^\text{fr}_{\pm_1,0}\over d\cos\theta}\sim {d\sigma^\text{fr}_{0,\pm_1}\over d\cos\theta} \sim {\alpha^4 \Lambda^2_{\rm QCD}\over  s^2},
\\
& & {d\sigma^\text{fr}_{0,0}\over d\cos\theta} \sim {d\sigma^\text{fr}_{\pm_1,\pm1}\over d\cos\theta} \sim {\alpha^4 \Lambda^4_{\rm QCD}\over  s^3},
\eqa
\eseq
where we assume $f_V\sim m_{V_i} \sim {\cal O}(\Lambda_{\rm QCD})$.
It is the $(\pm1,\mp1)$ helicity channel that yields the most dominant contribution.

Summing \eqref{polarized:differential:cross:section:vector} over all possible helicity configurations, we arrive at the fragmentation contribution to the
unpolarized cross section:
\begin{align}
   &{d\sigma^{\rm fr}(e^+ e^- \to V^0_1 V^0_2)\over d\cos\theta} = 16\pi^{3} \alpha^{4} \left( {{\cal Q}_{V_1} f_{V_1} \over m_{V_1}}\right)^2 \left( {{\cal Q}_{V_2} f_{V_2} \over m_{V_2}}\right)^2 \,\lambda^{1/2}(1,r_{V_1}^{2},r_{V_2}^2)
\nn \\
     & \times  { 2 s t u (m_{V_1}^2+m_{V_2}^2) + (t^2+u^2)(t u- m_{V_1}^{2} m_{V_2}^2)\over
     s\,t^2\,u^2}
\label{Equation:unpolarized:X:section:fragmentation}
\end{align}
where the Mandelstam variables $t$, $u$ are given by
\beq
    t =  -{s\over 2} (1-2r_{V_1}^2 -\sqrt{1-4r_{V_1}^2}\cos\theta),
\qquad
    u =  -{s\over 2} (1-2r_{V_2}^2 +\sqrt{1-4r_{V_2}^2}\cos\theta).
\eeq
Equation~\eqref{Equation:unpolarized:X:section:fragmentation} is in agreement with what is found in Refs.~\cite{Davier:2006fu,Bodwin:2006yd}.

\subsubsection{Interference part}
\label{interference}

With the knowledge of the fragmentation amplitude in \eqref{Frag:ampl:double:vector:meson} and the non-fragmentation amplitude in \eqref{helicity:amplitude:nonfrag:double:vector:meson}, we readily derive the interference part of the differential cross section. At leading-twist accuracy, the interference term becomes nonvanishing
only for the helicity configurations $(\lambda_1,\lambda_2)=(\pm, \mp)$ and $(0,0)$:
\bseq
\begin{align}
  \frac{d\sigma^\text{int}_{\pm1,\mp1}}{d\cos{\theta}}  =&-
{4\pi^3\alpha^4 \kappa_{V_1V_2} {\cal Q}_{V_1} {\cal Q}_{V_2} f_{V_1} f_{V_2} f_{V_1\perp} f_{V_2\perp}
         \over 3 r_{V_1}r_{V_2}s^3 }  \lambda^{1/2}(1,r_{V_1}^{2},r_{V_2}^{2})  (3+ \cos{2\theta})\, G_{\pm1,\mp1}(\cos\theta),
\\
 \frac{d\sigma^\text{int}_{0,0}}{d\cos{\theta}}
     =&-
  {32\pi^{3}\alpha^{4} \kappa_{V_1V_2} {\cal Q}_{V_1}{\cal Q}_{V_2} f_{V_1}^2 f_{V_2}^2 \over 3 s^{3}} \,\lambda^{1/2}(1,r_{V_1}^{2},r_{V_2}^{2}) \, G_{0,0}(\cos\theta),
\end{align}
\label{eq:sig:int}
\eseq
which bear the following asymptotical behaviors in the  limit $\sqrt{s}\to \infty$:
\beq
{d\sigma^\text{int}_{\pm_1,\mp_1}\over d\cos{\theta}} \sim  {\alpha^4 \Lambda^2_{\rm QCD}\over s^2},
\qquad
{d\sigma^\text{int}_{0,0}\over d\cos\theta} \sim {\alpha^4 \Lambda^4_{\rm QCD}\over  s^3}.
\label{scaling:interference:term}
\eeq
Therefore the dominant interference contribution is suppressed with respect to the dominant fragmentation contribution
by a factor of $\Lambda^2_{\rm QCD}/s$.

\subsubsection{Non-fragmentation part}
\label{vector:meson:nonfragmentation:part}

Squaring the non-fragmentation amplitude in \eqref{helicity:amplitude:nonfrag:double:vector:meson}, and averaging over the electron and positron spins,
we then obtain the non-fragmentation part of the differential cross section. For the same reason as in the interference part,
at the leading twist accuracy,
the non-fragmentation part survives only for the helicity configurations $(\lambda_1,\lambda_2)=(\pm, \mp), (0,0)$:
\bseq
\begin{align}
  \frac{d\sigma^\text{nf}_{\pm1,\mp1}}{d\cos{\theta}}
   =&\frac{\pi^{3}\alpha^{4} \kappa_{V_1V_2}^2 f_{V_1\perp}^2 f_{V_2\perp}^2 }
         {18 s^{3}} \sin^{2}{\theta}(3+\cos 2\theta) \,\lambda^{1/2}(1,r_{V_1}^{2},r_{V_2}^{2}) |G_{\pm1,\mp1}(\cos\theta)|^2 ,
\\
 \frac{d\sigma^\text{nf}_{0,0}}{d\cos{\theta}}
     =&\frac{\pi^{3}\alpha^{4}  \kappa_{V_1V_2}^2   f_{V_1}^2 f_{V_2}^2  }
         {9 s^{3}}\,\lambda^{1/2}(1,r_{V_1}^{2},r_{V_2}^{2})\,\sin^2\theta\,|G_{0,0}(\cos\theta)|^2,
\label{VV:00:polarized:X:section:non-fragmentation}
\end{align}
\label{eq:sig:nf}
\eseq
which possess the same asymptotical behaviors in the high energy limit:
\beq
{d\sigma^\text{nf}_{\pm_1,\mp_1}\over d\cos{\theta}} \sim  {\alpha^4 \Lambda^4_{\rm QCD}\over  s^3},
\qquad
{d\sigma^\text{nf}_{0,0}\over d\cos\theta} \sim {\alpha^4 \Lambda^4_{\rm QCD}\over  s^3}.
\eeq
Therefore the non-fragmentation part is suppressed with respect to the fragmentation contribution
by a factor of $(\Lambda_{\rm QCD}/s)^2$.

\subsection{Production rates for double neutral pseudoscalar mesons}

Since the exclusive production of a pair of neutral pseudoscalar pairs only proceeds through the non-fragmentation mechanism,
the step of deriving the corresponding differential cross section is identical to Sec.~\ref{vector:meson:nonfragmentation:part}.
Squaring the non-fragmentation amplitude in \eqref{amplitude:double:pseudoscalar:meson}, and averaging over the electron and positron spins,
we obtain
\begin{align}
 \frac{d\sigma^\text{nf}}{d\cos{\theta}}
 =&
 {\pi^{3}\alpha^{4} \kappa^2_{P_1P_2} f_{P_1}^2 f_{P_2}^2 \over 9s^{3}} \, \lambda^{1/2}(1,r_{V_1}^{2},r_{V_2}^{2})\,\sin^2\theta\, |G_{P,P}(\cos\theta)|^2.
\label{eq:sig:scal}
\end{align}
As mentioned before, the differential cross section for double pseudoscalar meson production very much resembles the non-fragmentation part of
the cross section for producing two longitudinally-polarized vector mesons, \eqref{VV:00:polarized:X:section:non-fragmentation},  which also scales as
$\alpha^4 \Lambda^4_{\rm QCD}/s^3$ in the limit $\sqrt{s}\to \infty$.  
This asymptotical scaling behavior is the same as that in the $\gamma\gamma \to \pi \pi$ process~\cite{Brodsky:1981rp}.

\section{Phenomenology}
\label{Phenomenology}

In this section, we present our predictions for the cross sections of various $e^+e^-\to M^0_1 M^0_2$ channels ($M=V,\,P$).
We consider two benchmark values of the $e^+e^-$ center-of-mass energy, $\sqrt{s}=3.77$ GeV for {\tt BESIII} experiment,
and $\sqrt{s}=10.58$ GeV for {\tt Belle} experiment.

\subsection{Input parameters}
\label{input:parameters}

\begin{table}[htbp]
  \begin{tabular}{|c|c|c|c|c|}
     \hline
 Vector                        &$\rho$          &$\omega$        &\multicolumn{2}{c|}{$\phi$}\\  \hline
 Mass\,[MeV]~\cite{Workman:2022ynf}                     &$775.26$&782.66&\multicolumn{2}{c|}{1019.46}\\ \hline
$f_{V}$[MeV]                   &$214.7\pm1.1$& $195.2\pm2.9$&\multicolumn{2}{c|}{$221.5\pm1.2$}\\
$f_{\perp}$[MeV]               &$165\pm9$~\cite{Ball:2007rt}& $151\pm9$~\cite{Ball:2007rt}&\multicolumn{2}{c|}{$186\pm9$~\cite{Ball:2007rt}}\\ \hline
$a_{1}^{\parallel(\perp)}$ ($\mu=2$GeV)&$0$             &$0$             &\multicolumn{2}{c|}{$0$}\\
$a_{2}^{\parallel}$ ($\mu=2$ GeV)&$\,\,0.132\pm0.027$~\cite{Braun:2016wnx}\,\,&$0.132\pm0.027$~\cite{Braun:2016wnx}&\multicolumn{2}{c|}{$0.13\pm0.06$\cite{Ball:2007rt}}\\
$a_{2}^{\perp}$ ($\mu=2$ GeV)&$\,\,0.101\pm0.022$~\cite{Braun:2016wnx}\,\,  &$0.101\pm0.022$~\cite{Braun:2016wnx}&\multicolumn{2}{c|}{$0.11\pm0.05$\cite{Ball:2007rt}}\\
     \hline\hline
Pseudoscalar        &$\pi^{0}$        &$K^{0}$        &$\eta_{q}$   &$\eta_{s}$ \\  \hline
 Mass\,[MeV]~\cite{Workman:2022ynf}          &134.98&497.61& 56.81 &707.01\\ \hline
$f_{P}$[MeV]        &$130.2\pm0.12$~\cite{Workman:2022ynf}&$155.7\pm0.3$~\cite{Workman:2022ynf}&$125\pm5$~\cite{Ottnad:2017bjt}&$178\pm4$~\cite{Ottnad:2017bjt}\\ \hline
$a_{1}$ ($\mu=2$GeV)&   0&  $-0.108\pm0.053$~\cite{LatticeParton:2022zqc}&0            &0\\
$a_{2}$ ($\mu=2$GeV)&$0.116^{+19}_{-20}$ ~\cite{RQCD:2019osh} &$0.170\pm0.046$~\cite{LatticeParton:2022zqc}&$0.116^{+19}_{-20}$ ~\cite{RQCD:2019osh}&$0.116^{+19}_{-20}$ ~\cite{RQCD:2019osh}\\
 $a_{3}$ ($\mu=2$GeV)& 0& $-0.043\pm0.023$~\cite{LatticeParton:2022zqc}& 0&0\\
 $a_{4}$ ($\mu=2$GeV)& $0.122\pm0.056$~\cite{LatticeParton:2022zqc}& $0.073\pm0.022$~\cite{LatticeParton:2022zqc}& $0.122\pm0.056$~\cite{LatticeParton:2022zqc}&$0.122\pm0.056$~\cite{LatticeParton:2022zqc}\\
$a_{6}$ ($\mu=2$GeV)& $0.068\pm0.038$~\cite{LatticeParton:2022zqc}& -& $0.068\pm0.038$~\cite{LatticeParton:2022zqc}&$0.068\pm0.038$~\cite{LatticeParton:2022zqc}\\  \hline
\end{tabular}
\caption{Masses, decay constants and a few lower-order Gegenbauer moments of the vector and pseudoscalar mesons. }
\label{tab:moments:num}
\end{table}

Our numerical predictions critically hinges on the profile of the LCDAs of various mesons,
which must be determined by the nonperturbative means.
As a common practice, one parameterizes the meson LCDA as the sum of
Gegenbauer polynomials:
\begin{eqnarray}
    \phi(x)=6x{\bar x}\Big[1+
    \sum_{n=1}^{\infty}a_{n}C^{3/2}_{n}(2x-1)\Big],
\end{eqnarray}
with $a_{n}$ representing the $n$-th Gegenbauer moment.

There have been extensive endeavors to unravel the LCDAs of the ground state mesons from different kinds of phenomenological approaches.
In recent years considerable progress has also been made on model-independent extractions of the LCDAs
from the lattice QCD simulation.  {\tt RQCD} collaboration has determined the $a_{2\parallel}$ and $a_{2\perp}$ moments for $\rho$~\cite{Braun:2016wnx},
as well as the $a_2$ moment for the $\pi$~\cite{RQCD:2019osh}. The {\tt LPC} collaboration has also predicted the whole profile of the pion and kaon LCDAs,
from which various Gegenbauer moments can be inferred~\cite{LatticeParton:2022zqc}. A very recent two-loop QCD analysis indicates that,
taking the {\tt RQCD} value of $a_2$, and the {\tt LPC} values of $a_4$ and $a_6$ can give a decent account for both the
time-like and space-like pion electromagnetic form factor data with large momentum transfer~\cite{Chen:2023byr}. We will adopt the same strategy
for the pion LCDA in the phenomenological analysis. Due to the lack of lattice simulation, we assume that the LCDAs of the
$\eta_q$ and $\eta_s$ are identical to that of the pion, yet differ in decay constants and masses~\cite{Feldmann:1998vh,Ali:2007ff}.
We take the values of the first four Gegenbauer moments of the $K^0$ meson
as given by {\tt LPC}~\cite{LatticeParton:2022zqc}. In the absence of the lattice result, we assume the Gegenbauer moments of $\omega$ is identical to those of $\rho$,
and take the moments of the $\phi$ meson from QCD sum rules analysis~\cite{Ball:2007rt}.

In addition, we take the values of the $\pi$ and $K^0$ decay constants from the latest compilation of PDG~\cite{Workman:2022ynf},
and take the lattice predictions of $\eta_q$ and $\eta_s$ decay constants given by
the {\tt ETM} collaboration~\cite{Ottnad:2017bjt}. We fix the values of the decay constants $f_V$ for ($V=\rho,\omega,\phi$)
from the measured leptonic widths~\cite{Workman:2022ynf}, and take the values of the
decay constants $f_{V\perp}$ from the QCD sum rules analysis~\cite{Ball:2007rt}.

For reader's convenience, in Table~\ref{tab:moments:num} we enumerate the values of the mass, decay constant and a few first Gegenbauer moments of light neutral
vector and pseudoscalar mesons.

In our numerical calculation, we choose to use the QED coupling constant $\alpha(\sqrt{s}/2)=1/133.46$ for {\tt BESIII}
and $\alpha(\sqrt{s}/2)=1/132.02$ for {\tt Belle} experiment.  These values of the running QED couplings are evaluated
by utilizing the package {\tt PYTHIA}~\cite{Bierlich:2022pfr}.

\subsection{Numerical results}
\label{Numerical:results}

\begin{table}[h]
    \centering
 \begin{tabular}{|c|c|c|c|c|} \hline
         &  \multicolumn{4}{|c|}{$\sqrt{s}=3.77$ GeV}\\ \hline
 process&  $\sigma^{\mathrm{fr}}$ [fb]&  $\sigma^{\mathrm{int}}$ [fb]&  $\sigma^{\mathrm{nfr}}$ [fb] & $\sigma^{\mathrm{tot}}$ [fb]\\\hline
         $e^{+}e^{-}\to\rho^{0}\rho^{0}$&   $622\pm 13$& $-66\pm8$& $3.2\pm0.6$& $560\pm15$\\ \hline
          $e^{+}e^{-}\to\rho^{0}\omega$&  $111.8\pm3.5$&  $-36.1\pm3.4$&  $5.3\pm0.7$& $71\pm5$\\ \hline
           $e^{+}e^{-}\to\rho^{0}\phi$&  $152.3\pm2.3$&  -&  - & $152.3\pm2.3$\\ \hline
 $e^{+}e^{-}\to \phi\phi$& $9.85\pm0.21$& $-4.4\pm0.7$& $0.74\pm0.13$& $6.2\pm0.7$\\\hline
 $e^{+}e^{-}\to\omega\phi$& $13.7\pm0.4$& -& - & $13.7\pm0.4$\\\hline
 $e^{+}e^{-}\to\omega\omega $& $5.02\pm0.30$& $-5.0\pm0.7$& $2.2\pm0.4$& $2.3\pm0.8$\\ \hline \hline
         &  \multicolumn{4}{|c|}{$\sqrt{s}=10.58$ GeV}\\ \hline
 process& $\sigma^{\mathrm{fr}}$ [fb]&  $\sigma^{\mathrm{int}}$ [fb]&  $\sigma^{\mathrm{nfr}}$ [fb] &$\sigma^{\mathrm{tot}}$ [fb]\\\hline
         $e^{+}e^{-}\to\rho^{0}\rho^{0}$&  $143.2\pm 2.9$& $-1.17\pm0.14$& $0.007\pm0.002$ &$142.0\pm2.9$\\ \hline
          $e^{+}e^{-}\to\rho^{0}\omega$&  $25.7\pm0.8$&  $-0.66\pm0.06$& $0.012\pm0.002$&$25.1\pm0.8$\\ \hline
           $e^{+}e^{-}\to\rho^{0}\phi$&  $36.1\pm0.5$&  -& - &$36.1\pm0.5$\\ \hline
 $e^{+}e^{-}\to \phi\phi$& $2.36\pm0.05$& $-0.08\pm0.01$&$0.002\pm0.0003$ &$2.27\pm0.05$\\\hline
 $e^{+}e^{-}\to\omega\phi$& $3.25\pm0.10$& -&- &$3.25\pm0.10$\\\hline
 $e^{+}e^{-}\to\omega\omega $& $1.16\pm0.07$& $-0.09\pm0.01$&$0.005\pm0.001$ &$1.07\pm0.07$\\\hline
    \end{tabular}
    \caption{Unpolarized  cross sections for $ e^{+}e^{-} \to V^0_{1} V^0_{2}$ at the benchmark {\tt BESIII} and {\tt Belle} energies.
    For each channel, we also enumerate the individual values of the fragmentation, interference  and the non-fragmentation parts.}
    \label{Table:X:Section:double:vector:meson:BES:BELLE}
\end{table}

In Table~ \ref{Table:X:Section:double:vector:meson:BES:BELLE} we enumerate the cross sections for various
exclusive double vector meson and double pseudoscalar production channels, both at {\tt BESIII} and {\tt Belle} energies.
For the case when the two vector mesons are identical particles, we have to integrate the differential cross section over only half of the entire phase space.

The total cross sections of three channels $e^+e^- \to \rho^0 \rho^0, \rho^0\omega, \rho^0\phi$ at $\sqrt{s}=3.77$ GeV are predicted to be $560\pm 15$, $71\pm5$, and $152.3\pm2.3$ fb,
respectively. To date the {\tt BESIII} experiment has accumulated about 20 fb$^{-1}$ data at this energy point, so there have already been
$10900\sim11500$, $1320\sim 1520$ and $3000\sim 3092$ events produced. Therefore, these three channels should be readily observed based on current {\tt BESIII} data set.

In contrast, the total production rates of the three processes $e^+e^+ \to \rho^0 \rho^0, \rho^0\omega, \rho^0\phi$ at $\sqrt{s}=10.58$ GeV are predicted to be $142.0\pm 2.9$, $25.1\pm 0.8$ and $36.1\pm 0.5$ fb, respectively.
The cross sections are more than three times smaller than those at {\tt BESIII}, which is compatible with the asymptotic scaling behavior $\sigma^\text{fr}(\sqrt{s}) \sim  \alpha^4/s$ in
\eqref{Fragmentation:cross:section:Leading:helicity:scaling},  since the double neutral vector meson production is dominated by the fragmentation mechanism at high energy.

Thus far the integrated luminosity of {\tt Belle} experiment has reached about 1500 fb$^{-1}$, and we expect about $208650\sim 217350$ $\rho^0 \rho^0$ events,
$36450\sim 38850$  $\rho^0\omega$ events, and $53400\sim 54900$ $\rho^0\phi$ events have been produced.
 These double vector meson events are so copious and deserve a dedicated study experimentally.

Equation~\eqref{Equation:unpolarized:X:section:fragmentation} indicates that the fragmentation contributions to the three production channels obey the hierarchy:
$\sigma(\rho^0 \rho^0): \sigma(\rho^0 \omega): \sigma(\rho^0 \phi)\approx {\cal Q}^2_{\rho}/2: {\cal Q}^2_\omega: {\cal Q}^2_\phi = 9: 2: 4$, where the
effective quark electric charge in a vector meson given in \eqref{effective:quark:charge}~\footnote{Owing to the smaller effective quark charge in $\omega$ and $\phi$ relative to
that in $\rho^0$, the production rates for  $e^+e^- \to \phi\phi, \phi\omega, \omega\omega$ are orders of magnitude suppressed with respect to those for $e^+e^-\to \rho^0\rho^0, \rho^0\omega, \rho^0\phi$.}.  We clearly see from
Table~\ref{Table:X:Section:double:vector:meson:BES:BELLE} that the predicted cross sections at {\tt Belle} fit into this pattern better than those at {\tt BESIII},
since the fragmentation dominance is a much better approximation in the higher energy $e^+e^-$ collider.

In addition to the total cross section, in Table~\ref{Table:X:Section:double:vector:meson:BES:BELLE} we also enumerate the individual contribution from the fragmentation,
interference and non-fragmentation parts for a variety of double vector meson production channels. 
One clearly sees that the interference contributions are destructive. For the {\tt Belle}, the magnitude of the interference piece is smaller than the uncertainty of
the fragmentation contribution. Since the estimated fragmentation contribution~\cite{Davier:2006fu,Bodwin:2006yd} already 
successfully accounts for the measured production rates of the
$e^+e^-\to \rho^{0}\rho^{0}$ and $\rho^0\phi$ processes by {\tt BaBar}~\cite{BaBar:2006vxk}, it seems unfeasible to 
unambiguously pinpoint the interference contribution from the experimental perspective, even with a great amount of events at {\tt Belle} experiment.

As can be clearly visualized in Table~\ref{Table:X:Section:double:vector:meson:BES:BELLE}, 
it turns out that the interference effect is much more significant at {\tt BESIII} experiment. For the $e^+e^-\to \rho^{0}\rho^{0}$ channel, including the interference
contribution can lower the fragmentation prediction by about 10\%; For the $e^+e^-\to \rho^{0}\omega$ channel, including the interference contribution may even lower
the fragmentation contribution by 30\%.  
It is easy to understand why the interference contribution becomes notable at {\tt BESIII} but negligible at {\tt Belle}, since  
the interference contribution is suppressed with respect to the fragmentation contribution
by a factor of $\Lambda^2_{\rm QCD}/s$, as indicated by \eqref{Fragmentation:cross:section:Leading:helicity:scaling} and \eqref{scaling:interference:term}.

\begin{figure}[htb]
\includegraphics[width=1\textwidth]{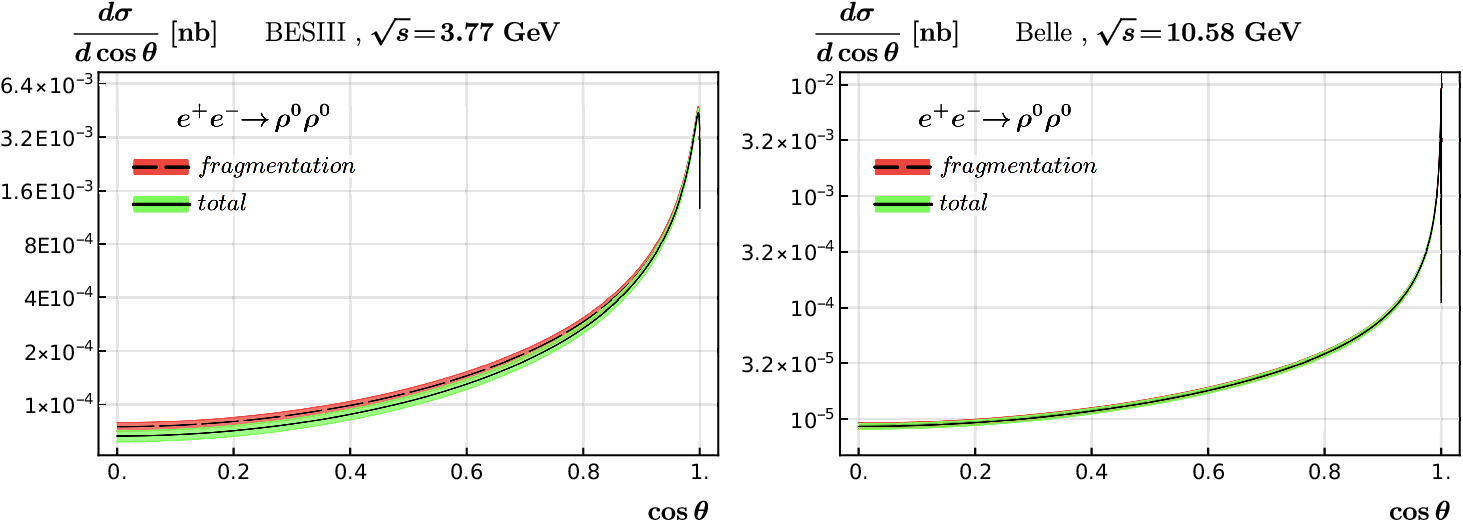}
 \caption{The angular distribution of the $\rho^0$ in the process $e^{+}e^{-} \to \rho^{0}\rho^{0}$.
 The red band represents the contribution from fragmentation mechanism only, while the green band represents the sum of the
 fragmentation, interference, and non-fragmentation contributions.}
\label{Fig:rho:rho:angular:cross section:distribution}
\end{figure}

\begin{figure}[htb]
\includegraphics[width=1\textwidth]{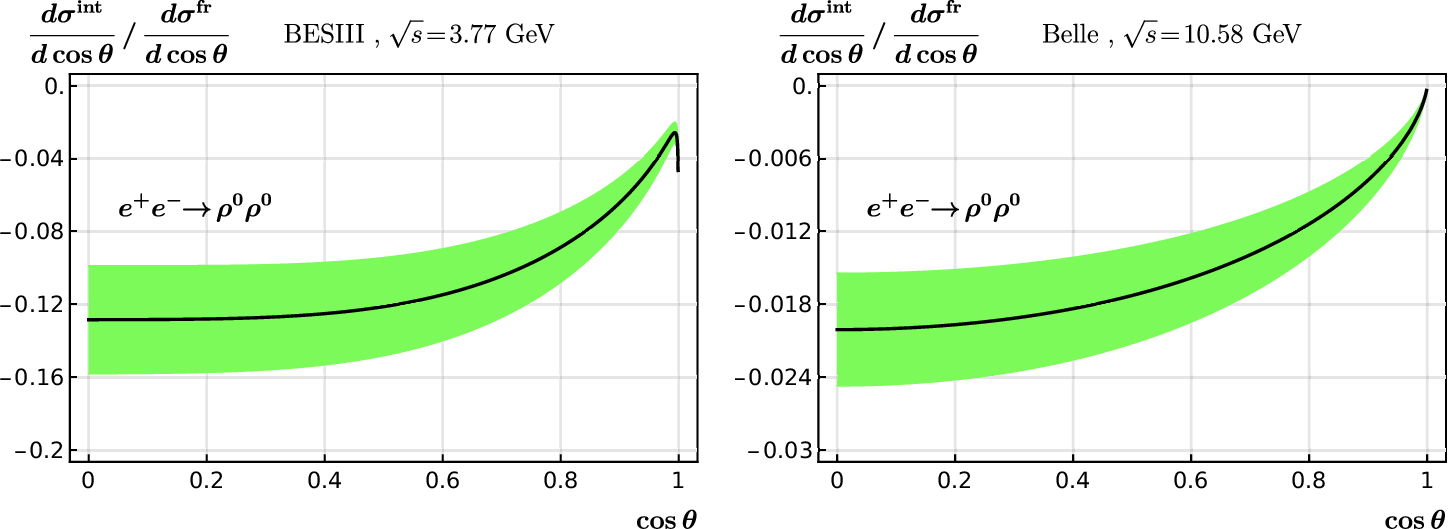}
\caption{The ratio of the interference part to the fragmentation part in the process $e^{+}e^{-} \to \rho^{0}\rho^{0}$ as a function of $\cos\theta$.}
\label{ratio:interference:to:frag:rho:rho}
\end{figure}

\begin{figure}[htb]
\includegraphics[width=1\textwidth]{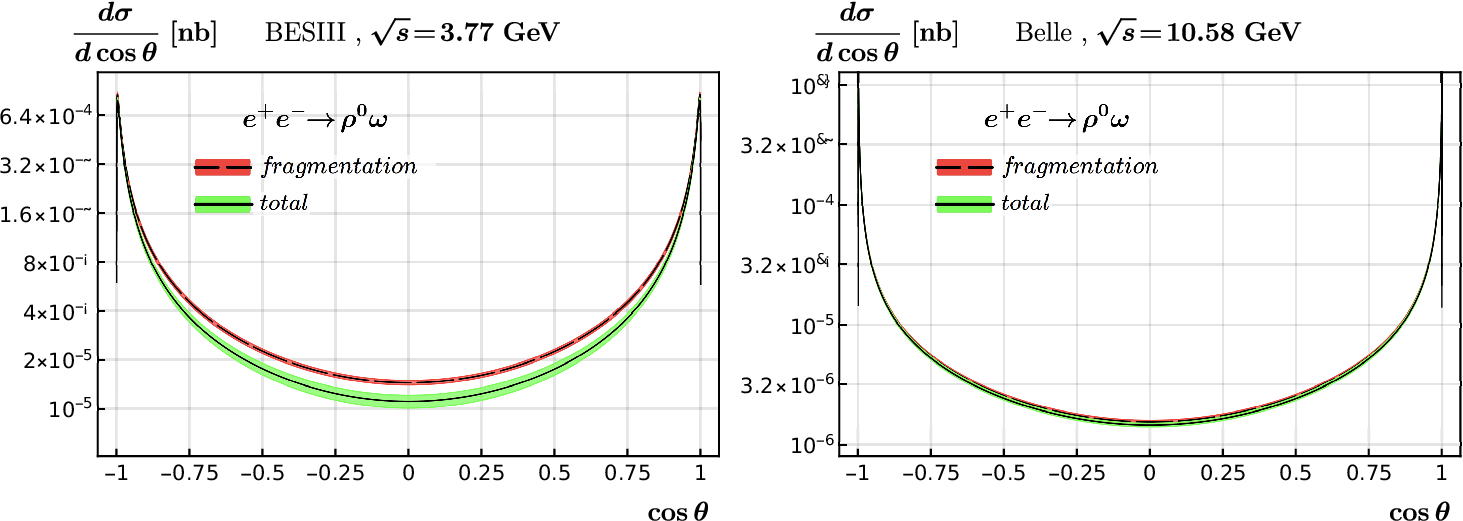}
 \caption{The angular distribution of the $\rho^0$ in the process $e^{+}e^{-} \to \rho^{0}\omega$. The red band represents the fragmentation
 contribution, while the green band represents the sum of three pieces of contributions.}
\label{Fig:rho:omega:angular:cross section:distribution}
\end{figure}

\begin{figure}[htb]
\includegraphics[width=1\textwidth]{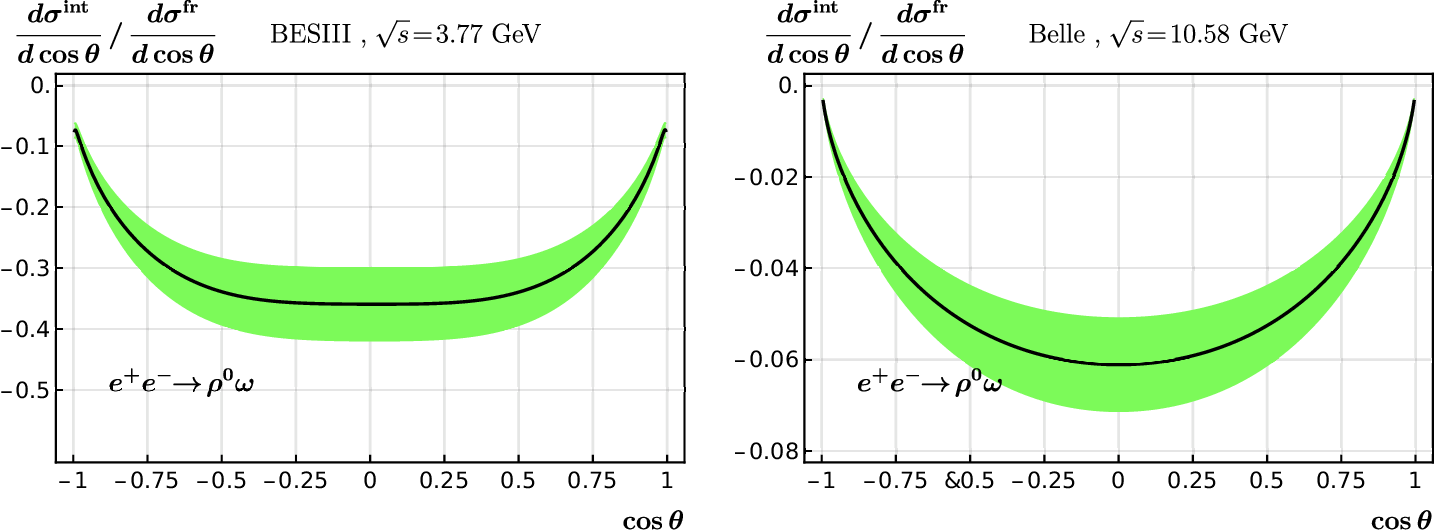}
\caption{The angular distribution of the ratio of the interference part to the fragmentation part in the process $e^{+}e^{-} \to \rho^{0}\omega^{0}$.}
\label{ratio:interference:to:frag:rho:omega}
\end{figure}

The impact of the interference contribution for $e^+e^-\to \rho^0\rho^0$ can also been seen in the angular distribution of $\rho^0$ in 
Fig.~\ref{Fig:rho:rho:angular:cross section:distribution} and Fig.~\ref{ratio:interference:to:frag:rho:rho}. 
Clearly, the interference effect in this channel becomes already notable at {\tt BESIII} experiment. 
More interestingly, as can be seen from the angular distribution of the $\rho^0$ meson in 
Fig.~\ref{Fig:rho:omega:angular:cross section:distribution} and Fig.~\ref{ratio:interference:to:frag:rho:omega},
the impact of the interference contribution the for $e^+e^-\to \rho^0\omega$ channel becomes significant at 
{\tt BESIII} energy.

We also observe from Table~\ref{Table:X:Section:double:vector:meson:BES:BELLE} that, the non-fragmentation contribution in various double vector meson production processes, 
which receive a $\Lambda^4_{\rm QCD}/s^2$ suppression with respect to the fragmentation prediction, become completely negligible, 
even at {\tt BESIII} energy~\footnote{An exception is the $e^+e^-\to \omega\omega$ channel at {\tt BESIII} energy, 
where the fragmentation and interference pieces coincidentally nearly cancel with each other,
so the non-fragmentation contribution becomes the dominant one. }.  

We thus urge the future {\tt BESIII} experiment to conduct a precise measurement of the differential cross sections for 
the $e^+e^-\to \rho^{0}\rho^{0}$ and $e^+e^-\to \rho^{0}\omega$ channels. 
Once the definite deviation from the fragmentation prediction is observed, 
one may unambiguously establish the existence of the destructive interference contribution.
By confronting theoretical predictions and experimental measurements, our knowledge about the LCDAs of $\rho^0$ and $\omega$
will be greatly enhanced.

\begin{table}[h]
    \centering
    \begin{tabular}{|c|c|c|c|c|c|c|} \hline
         &  \multicolumn{3}{|c|}{{\tt BESIII}, $\sigma^{\mathrm{tot}}~[\mathrm{ab}]$}&  \multicolumn{3}{|c|}{{\tt Belle}, $\sigma^{\mathrm{tot}}~[\mathrm{ab}]$}\\\hline \hline
          $e^{+}e^{-}\to\pi^{0}\pi^{0}$&  \multicolumn{3}{|c|}{$444\pm212$}&  \multicolumn{3}{|c|}{$1.0\pm0.5$}\\ \hline
          $e^{+}e^{-}\to\pi^{0}\eta$&  \multicolumn{3}{|c|}{$175\pm85$}&  \multicolumn{3}{|c|}{$0.38\pm0.5$}\\ \hline
           $e^{+}e^{-}\to\pi^{0}\eta^{'}$&  \multicolumn{3}{|c|}{$108\pm53$}&  \multicolumn{3}{|c|}{$0.25\pm0.12$}\\ \hline
 $e^{+}e^{-}\to\eta\eta$& \multicolumn{3}{|c|}{$310\pm152$}& \multicolumn{3}{|c|}{$0.69\pm0.34$}\\\hline
 $e^{+}e^{-}\to\eta^{'}\eta^{'}$& \multicolumn{3}{|c|}{$255\pm124$}& \multicolumn{3}{|c|}{$0.62\pm0.30$}\\\hline
 $e^{+}e^{-}\to\eta \eta^{'}$& \multicolumn{3}{|c|}{$2.9\pm3.1$}& \multicolumn{3}{|c|}{$0.007\pm0.007$}\\\hline
 $e^{+}e^{-}\to K^{0}_{s}K^{0}_{s}$& \multicolumn{3}{|c|}{$121\pm35$}& \multicolumn{3}{|c|}{$0.27\pm0.08$}\\\hline
    \end{tabular}
    \caption{Total cross sections for $ e^{+}e^{-} \to P^0_{1} P^0_{2}$ at {\tt BESIII} and {\tt Belle} energies.}
    \label{Table:X:Section:double:pseudoscalar:meson:BES:BELLE}
\end{table}

\begin{figure}[htb]
\includegraphics[width=1\textwidth]{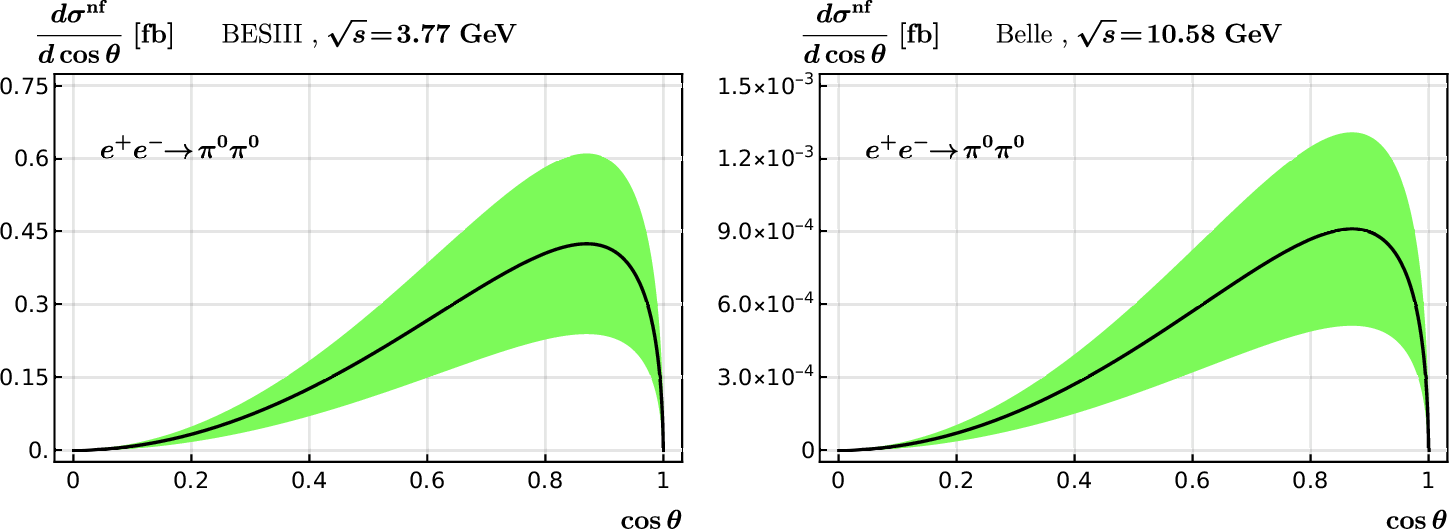}
\caption{Angular distribution of the $\pi^0$ in the $e^{+}e^{-} \to \pi^{0}+\pi^{0}$ process.}
\label{Fig:pi:omega:angular:cross section:distribution}
\end{figure}

In Table~\ref{Table:X:Section:double:pseudoscalar:meson:BES:BELLE} we also enumerate the cross sections for various
exclusive double neutral pseudoscalar meson production channels, both at {\tt BESIII} and {\tt Belle} energies.
The production rates turn out to be exceedingly tiny, which reach only a few tenth of fb for the 
$e^+e^-\to \pi^0\pi^0, \eta\eta, \eta'\eta', \pi^0\eta$ channels at $\sqrt{s}=3.77$ GeV. For concreteness, 
we plot the angular distribution of the $\pi^0$ for the $e^{+}e^{-} \to \pi^{0}+\pi^{0}$ process in
Fig.~\ref{Fig:pi:omega:angular:cross section:distribution}.
The smallness of the cross sections is clearly rooted in the $\alpha^4 \Lambda^4_{\rm QCD}/s^3$ scaling behavior 
dictated by the non-fragmentation mechanism. 
It looks rather challenging to observe these rare exclusive production channels at {\tt BESIII} experiment.

\section{Summary}
\label{Summary}

In this work, we have conducted a comprehensive investigation on the exclusive production of double light neutral mesons (with $C=+1$) from $e^+e^-$ annihilation, 
exemplified by the {\tt BESIII} and {\tt Belle} experiments. 
These exclusive processes necessarily entail two-photon exchange, with the double neutral meson production mechanism 
classified into the photon fragmentation and non-fragmentation.
The fragmentation contribution is only relevant for double neutral vector meson production, 
which can be reliably estimated in a rigorous manner, by taking the measured leptonic width of the neutral vector meson as input. 
The non-fragmentation amplitudes are computed within the framework of collinear factorization, 
to the lowest order in QED and QCD coupling and at leading twist accuracy. 

For the exclusive production of a pair of light neutral pseudoscalar mesons, only the non-fragmentation production mechanism survives,
and the corresponding production rates are too small for these types of processes to be observed at both {\tt BESIII} and {\tt Belle}.
The exclusive production of a pair of light neutral vector mesons is more interesting. The cross sections are generally quite large owning to 
the kinematic enhancement brought by the fragmentation mechanism, so that such processes should be readily observed and precisely measured in
both {\tt BESIII} and {\tt Belle} experiments. 
Although the fragmentation contribution plays the dominant role, including the non-fragmentation contribution may bring in
a sizable destructive interference effect at {\tt BESIII} energy.  We find that, with certain choice of the $\rho^0$/$\omega$ LCDAs 
inspired by the lattice QCD study, including the interference effect may lower the fragmentation contribution to the production rate for
$e^+e^-\to \rho^0\rho^0$  at $\sqrt{s}=3.77$ GeV by about $10\%$, and decrease the fragmentation prediction for $e^+e^-\to \rho^0\omega$
by about $30\%$. Thanks to a copious number of $\rho^0\rho^0$ and $\rho^0\omega$ events produced in the {\tt BESIII} experiment,
it is feasible to accurately measure the angular distributions of the these processes. 
Observation of explicit deviation from the fragmentation predictions
will unambiguously indicate the existence of the interference effect. 
Confronting the future {\tt BESIII} measurements with our predictions, it seems possible to impose useful constraints on the profile of the $\rho^0$/$\omega$ LCDAs.
The new knowledge gleaned from the double vector meson production at {\tt BESIII} 
may in turn be beneficial to precise predictions of $B\to V$ form factor and $B\to VV$ decay rates.

\begin{acknowledgments}
{\noindent\it Acknowledgment.}
We thank Deshan Yang for discussions. The work of C.-P.~J., Y.~J. and J.-L.~L is supported in part by the National Natural Science Foundation of China
under Grants No.~11925506.
The work of X.-N.~X. is supported in part by the National Natural Science Foundation of China under Grant No.~12275364.
\end{acknowledgments}

\end{document}